\documentclass[12pt]{article}
\usepackage{amssymb}
\usepackage{hyperref}
\usepackage{epsfig}
\newtheorem{theorem}{\sc Theorem}[section]
\newtheorem{definition}{\sc Definition}[section]
\newtheorem{lemma}{\sc Lemma}[section]

\newtheorem{example}{\sc Example}[section]
\newtheorem{remark}{\it Remark}[section]
\newtheorem{Ack}{\it Acknowledgements}[section]
\begin{document}
\baselineskip=24 pt 
\title{\bf   The discontinuous dynamics and non-autonomous chaos}

\author{ M. U. Akhmet}

\date{{\small Department of Mathematics and Institute of Applied Mathematics, Middle East
Technical University, 06531 Ankara, Turkey}}
\maketitle

\noindent {\bf Address:} M. Akhmet,
Department of Mathematics, Middle East
Technical University, 06531 Ankara, Turkey,\\
{\bf fax:} 90-312-210-12-82\\
\noindent {\bf e-mail:} marat@metu.edu.tr
 
\vspace{0.5cm}

\newpage
\begin{abstract}
A multidimensional  chaos is generated by a special initial  value problem for the  non-autonomous impulsive differential equation.  The existence of a chaotic attractor is shown, where  density of periodic solutions, sensitivity of solutions and existence of a trajectory dense in the set of all  orbits  are observed. The chaotic properties of all solutions are discussed. An appropriate example is  constructed, where the intermittency phenomenon is indicated.  The results of the paper are illustrating  that impulsive differential equations  may play a special role in the investigation of the complex behavior of dynamical systems, different from that played by  continuous dynamics.
\end{abstract}
\maketitle

\section{Introduction and Preliminaries}

       The investigation of the complex behavior of dynamical systems, as well as the development of the methods of this  investigation,   has made a long way, starting with  significant  works \cite{birkhoff}- \cite{ueda}.
       
         It is   natural {\it to discover a chaos} \,  \cite{ birkhoff}-\cite{chua},\cite{henon}, \cite{lorenz}--\cite{ruelle1},\cite{smale,ueda},  and proceed by producing  basic definitions and creating  the theory. On the other hand, one can,  following the prescriptions, {\it shape}  an irregular process by inserting  chaotic elements in a system which has  regular dynamics otherwise  (let's say  is asymptotically stable, has a global attractor, etc). This  approach to the problem also deserves  consideration as it may allow for  a more rigorous treatment  of the  phenomenon, and help develop new  methods of investigation.
         Our results are of this  type. Using the logistic map as a generator of moments of  impulses in the  multidimensional system, we observe the density of the periodic solutions, the sensitivity of solutions and the existence of a trajectory dense in the set of all  solutions in a bounded region of the space. 
                
         The first mathematical definition of chaos was introduced by Li and Yorke \cite{li}. They proved that if a map on an interval had a point of period three, then it had points of all periods. Moreover, there exists an uncountable {\it scrambled} subset of the interval. Devaney  \cite{dev} gave an explicit definition of a chaotic invariant set in an attempt to clarify the notion of chaos. To the properties of {\it transitivity} and {\it sensitivity} \cite{ruelle1} he added the assumption that the periodic points are dense in the space.  An intensive discussion of the definitions has continued during the last decades. It was shown in \cite{kuchta} that a map is Li-Yorke chaotic if and only if there is a two-point scrambled set.  Paper  \cite{huang} proves that chaos as defined by Devaney is stronger than that defined by  Li-Yorke. In   \cite{banks} it was shown that transitivity and density of periodic points  imply sensitivity. 
         
 In this paper we concentrate on the  topological components of the version proposed by  Devaney.       
     The  special initial value problem  is introduced, when the moments of the impulsive action are functionally dependent on the initial moment.

 One of the most powerful tools of the chaos investigation is the conjugacy with the symbolic dynamics. Our results are based on the method, too. Consider the sequence space  \cite{dev}
	\[\Sigma_2 = \{s = (s_0s_1s_2\ldots): s_j = 0\,\, {\mbox or}\,\, 1\}
\]
with the metric 
	\[d[s,t] = \sum\limits_{i=0}^{\infty}\frac{|s_i-t_i|}{2^i},
\]
where $t=(t_0t_1\ldots) \in \Sigma_2,$ and the shift map $\sigma:  \Sigma_2 \to \Sigma_2,$ such that
$\sigma(s) = (s_1s_2\ldots).$  The semidynamics $(\Sigma_2,\sigma)$ is  the symbolic dynamics \cite{wig}.    

The map is continuous, $card Per_n(\sigma) = 2^n, Per(\sigma)$ is dense in $\Sigma_2,$ and there exists a dense orbit in $\Sigma_2.$

We assume that there exist  a homeomorphism $S$ between $\Sigma_2$ and a set $\Lambda \subset I, I = [0,\bar \omega],$ where $\bar \omega$ is a fixed positive number, and   a map $h: \Lambda \to \Lambda,$ such that $S \circ h = \sigma \circ S.$ That is $h$ and $\sigma$ are topologically conjugate. It is known that  $\Sigma_2$ and  $\Lambda$ are Cantor sets: they are closed, perfect and totally disconnected \cite{wig}. Obviously, they are compact. Moreover,  $\sigma$ and $h$ have sensitive dependence on the initial conditions,  periodic points are dense in $\Sigma_2$ and $\Lambda$ respectively,  and the maps are  topologically transitive.  We may  also specify  that for each $p \in \mathbb N$ there exists a solution with period $p,$ and  topological transitivity  means having  a  dense trajectory. 

One of the most popular examples of the map $h$ is the logistic map 
 $\mu x^2(1-x), \mu >4,$ considered on  a subset of $[0,1],$ \cite{rob}.

The  description of the main subject  of our paper  should begin  with the discussion of the  moments of  impulses, as their generation is  most important for the emergence of  chaos.

For every  $t_0 \in \Lambda $ one can construct a sequence $\kappa(t_0)$ of real numbers $\kappa_i, i \in \mathbb Z,$ in the following way. If $i \ge 0,$ then  $\kappa_{i+1} = h(\kappa_{i})$ and $\kappa_0 = t_0.$  
Let us show, how the sequence is defined for negative $i.$
 Denote $s^0 = S(t_0), s^0 = (s^0_0s^0_1\ldots).$ Consider elements $\underline{s} =   (0s^0_0s^1_0\ldots),\overline{s} =   (1s^0_0s^1_0\ldots)$ of\, $\Sigma_2,$ such that $\sigma(\underline{s})= \sigma(\overline{s})= s^0$ and 
$\underline{t} = S^{-1}(\underline{s}), \overline{t} = S^{-1}(\overline{s}).$ The homeomorphism implies that
$h( \overline{t}) = h( \underline{t}) = t_0.$  Set $h^{-1}(t_0)$ may consist of not more than two elements  $\overline{t},\underline{t} \in \Lambda.$ Each of these two values can be chosen as $\kappa_{-1}(t_0).$  Obviously, one can continue the process to $-\infty,$ choosing always one element from the set $h^{-1}.$ We have finalized the construction of the  sequence, and, moreover, it is proved that $\kappa(t_0) \subset \Lambda.$ 
Thus, infinitely many sequences  $\kappa(t_0)$   can be constructed for a given $t_0.$ However, each of this type of sequence  is  unique for an increasing $i.$ Fix one of the sequences  and  define a sequence  $\zeta(t_0) = \{\zeta_i\},  \zeta_i = i\bar \omega + \kappa_i, i \in \mathbb Z.$
The sequence has a  {\it periodicity property} if there exists  $p \in \mathbb N$ such that $\zeta_{i+p} = \zeta_i + p\bar\omega, \forall i \in \mathbb Z.$
If we denote by  $\Pi$ the set of  all such sequences $\{\zeta_i\}, i \in \mathbb Z,$ then  a multivalued functional
$w: I \to \Pi$ such that $ \zeta(t_0)=w(t_0)$ is defined. 

In what follows, we assume, without loss of generality, that $\bar \omega = 1.$

 Let $J \subseteq \mathbb R$  be an open interval. We introduce the distance $\|\zeta(t_0)-\zeta(t_1)\|_J = \sup\limits_{\zeta_i(t_0),\zeta_i(t_1) \in J}|\zeta_i(t_0)-\zeta_i(t_1)|,$ and, if $\zeta(t) = \{\zeta_i(t)\}$ is a sequence from $\Pi,$ and $m$ is an integer, we denote $\zeta(t,m) = \{\zeta_{i+m}(t)\}, i \in \mathbb N.$ More information about the opportunities  connected with this distance can be found in \cite{ak2}.

Next, let us consider some  useful properties of the elements of $\Pi.$ They are simple  consequences of the topological conjugacy.  
The first one is a reformulation of the  known  property  for  symbolic dynamics \cite{wig}.
\begin{lemma}\label{lempi} Assume that $h$ is topologically conjugate to $\sigma$ then the following assertions are valid:  
\begin{enumerate}
	\item[(a)] for each $\zeta(t_0) \in \Pi,$  an arbitrary small $\epsilon > 0,$ and an arbitrary large positive number $E$ there exists a  sequence $\zeta(t_1) \in \Pi$ with the periodicity property such that 
	$|\zeta(t_0)-\zeta(t_1)|_J < \epsilon,$ where $J = (0,E)$ 
		\item[(b)] There exists a sequence 
		$\zeta(t^*) \in \Pi$ such that  for each $t_0 \in \Lambda,$ and an arbitrary small $\epsilon > 0,$ and an arbitrary large positive number $E$ there exists  an integer $m$ such that 	$|\zeta(t_0)-\zeta(t^*,m)|_J < \epsilon,$ where $J = (0,E).$
	
\end{enumerate}
\end{lemma} 

\textbf{Proof:} $(a)$ Fix numbers $t_0 \in \Lambda, \epsilon >0, E >0,$ and denote $S(t_0) = s^0 = (s_0^0s_1^0\ldots).$
Since $S$ is a homeomorphism and $\Sigma_2$ is compact, there exists a number $\delta >0$ such that $|S^{-1}(s^1) - S^{-1}(s^2)| < \epsilon$ if $d[s^1,s^2] < \delta,$ where $s^1,s^2 \in \Sigma_2.$

Next we define a periodic sequence $s = (s_0s_1s_3\ldots)$ from  $\Sigma_2$ to satisfy $\epsilon$ and $E.$ Take a number $l \in \mathbb N$ such that $l > E$ and $\frac{1}{2^{l-1}} < \delta.$  Assume that $s_i = s_i^0, i = 0,1,\ldots,2l-1,$ and $s_{i+2l} = s_i, i \in \mathbb Z.$ Consider the sequence $\kappa_i(t_0) = h^i(t_0), i \ge 0.$ Since $h^i(t_0) = S^{-1} \circ \sigma^i \circ S,$ and $d[\sigma^i s^0, \sigma^i s] < \delta, i = 0,1,\ldots,l,$ we have that 
$|h^i(t_0) - h^i(S^{-1}(s)| < \epsilon.$ So, if we denote $t = S^{-1}(s)$ then $|\kappa_i(t_0) - \kappa_i(t)| < \epsilon, i =   0,1,\ldots,l.$ In other words, $\|\zeta(t_0) - \zeta(t)\|_J < \epsilon,$ where $J= (0,E).$ The assertion is proved.

$(b).$ Consider the sequence $s^* \in \Sigma_2$ such that 
	\[s^* = \underbrace{01}_{ 1 \, element \, blocks}|\underbrace{00011011}_{ 2\,element \, blocks}|\ldots,
\]
that is $s^*$ is constructed by successively listing all blocks of $0's$ and $1's$ of length $1,$ then length $2,$ etc. Sequence $s^*$ is dense in  $\Sigma_2$ \, \cite{dev}. Denote $t^* = S^{-1}(s^*).$  Let us fix $t_0 \in \Lambda, \epsilon > 0$ and $E >0.$ Similarly  to the previous proof, fix a number $\delta >0$ such that $|S^{-1}(s^1) - S^{-1}(s^2)| < \epsilon$ if $d[s^1,s^2] < \delta,$ where $s^1,s^2 \in \Sigma_2.$
Take $l \in \mathbb N$ such that $l > E+1$ and $\frac{1}{2^{l-1}} < \delta.$ 
We can find  $2l-$ block $s_m^*s_{m+1}^*\ldots s_{m+2l}^*$ such that $i-$th element 
of $S(t_0), i =0,1,\ldots,2l,$ is equal to $s_{i+m}^*.$   One can see that $\|\zeta(t_0) - \zeta(t^*,m)\|_J < \epsilon,$ where $J = (0,E).$ 
The lemma is proved.




 The following special initial value problem for the   impulsive  differential equation, 
\begin{eqnarray}
	&&z'(t) = A z(t) + f(z), t \not = \zeta_i(t_0),\nonumber\\
	&&\Delta|_{t = \zeta_i(t_0)} = B z(\zeta_i(t_0)) + W(z(\zeta_i(t_0))),\nonumber\\
	&&z(t_0) = z_0,\, (t_0, z_0)  \in \Lambda \times \mathbb R^n,
	\label{1}
\end{eqnarray}
where $z \in \mathbb R^n, t \in \mathbb R,$ and  $ \zeta(t_0)=w(t_0),$ 
is the object, which  will be mainly discussed in our paper. 

We shall need the  following  basic assumptions for the system:

 \begin{itemize}
\item[(C1)]  $A,B$ are $n\times n$ constant real valued matrices, $\det({\cal I} +B) \not = 0,$ where ${\cal I}$ is 
$n\times n$ identical matrix;
\item[(C2)] for all $x_1,x_2 \in \mathbb R^n$ functions $f(x,y):\mathbb R^n\times  \mathbb R^n  \to \mathbb R^n, W:\mathbb R^n  \to \mathbb R^n ,$  
 satisfy
\begin{eqnarray}
	&& ||f(x_1)-f(x_2)|| +\|W(x_1) -W(x_2)\| \le L||x_1 - x_2||,
\label{top-top}  
\end{eqnarray}
where $L > 0$ is a constant;
\item[(C3)] 
\begin{eqnarray}
	&&  Bx + W(x) \not = 0, \forall x \in \mathbb R^n;
\label{tip-tip}  
\end{eqnarray}
\item[(C4)]  functions $f$ and $W$ are uniformly bounded so that
 \begin{eqnarray}
	&& \sup_{x \in \mathbb R^n}||f(x)|| +  \sup_{x \in \mathbb R^n}||W(x)|| = M_0 < \infty.
\label{tyap-tup}  
\end{eqnarray}

\end{itemize}

We considered the systems of  form (\ref{1}) in \cite{akh23} in the investigation of  blood pressure distribution.

In the present  paper the sensitivity, transitivity and  existence  of dense   periodic solutions of  nonlinear problem (\ref{1}) are considered.  We also prove the  existence of the chaotic attractor, which consists of all solutions bounded on the real axis and discuss the  period-doubling cascade for the problem.


All the  results are rigorously verified using the  results of the theory of impulsive differential equations \cite{hw}-\cite{a4}.  

While  many theorems  of the theory of impulsive differential equations are obtained by analogy with the  assertions which already exist for the systems with continuous solutions, the results of the present work  do not have direct predecessors from the continuous dynamics.

The investigation  is  inspired by the discontinuous dynamics  of  the neural information processing in the brain, information communication, and population dynamics  \cite{glass}-\cite{wang}. While there are many interesting papers 
concerned with the complex behavior generated by impulses,  the rigorous theory of chaotic impulsive systems remains far from being  complete. Our goal is to develop further the theoretical foundations of this area of research.

The paper is organized in the following manner. The existence of the attractor is under discussion  in the next section.  In section 3 we consider the main subject of the paper: ingredients of chaos, the chaotic attractor,  the period-doubling cascade  and an example.
 
 \section{An attractor consisting of the solutions \\bounded on $\mathbb R$}
 
 In this section we discuss the existence of bounded and periodic solutions of the initial value problem, and the existence of an  attractor consisting  of the bounded on $\mathbb R$ solutions.

 Let us start with a system of a more general form than (\ref{1}).
 
 Fix a  sequence $\theta = \{\theta_i\}, i \in \mathbb Z,$   such that  
 $ i  \le \theta_i < i+1, i \in \mathbb Z,$ 
 and consider the following initial value problem  
\begin{eqnarray}
	&&z'(t) = A z(t) + f(z), t \not = \theta_i,\nonumber\\
	&&\Delta|_{t = \theta_i} = B z(\theta_i) + W(z(\theta_i)),\nonumber\\
	&&z(\theta_0) = z_0, z_0 \in \mathbb R^n,
\label{1+}
\end{eqnarray}
 where matrices $A$ and $B,$ functions $f$ and $W$ are the same as in (\ref{1}).
 
 In order to define the solutions of the impulsive systems we need the following spaces of discontinuous functions.
 
 A left continuous function $z(t)$ is from a set of functions ${\cal PC}^1(\theta,\mathbb R),$ if:
\begin{enumerate}
\item[(i)] it has discontinuities only  at points $\theta_i, i \in \mathbb Z,$ and they are of the first kind; 
\item[(ii)] the derivative $z'(t)$ exists at each point $t \in \mathbb R\backslash \{\theta_i\},$ and 
at points $\theta_i, i \in \mathbb Z,$ the left-sided derivative exists.
\end{enumerate}

A function $z(t) \in {\cal PC}^1(\theta,\mathbb R)$ is a solution of (\ref{1+}) if:
\begin{enumerate}
\item[(1)] the differential equation  is satisfied for $z(t)$ on $\mathbb R\backslash \{\theta_i\},$ and it holds for the  left derivative of $z(t)$ at every point $\theta_i, i \in \mathbb Z.$ 
\item[(2)] the jumps equation  is satisfied by $z(t)$ for every   $ i \in \mathbb Z.$ 
\end{enumerate}
It is known \cite{lbs,samo} that under the above mentioned conditions the solution of (\ref{1+}) exists, is unique on $\mathbb R,$ and 
it is an  element of  ${\cal PC}^1(\theta,\mathbb R).$

We shall also need the following set of functions.

 A left continuous function $z(t):[\theta_0,\infty)  \to \mathbb R^n$ is from a set of functions ${\cal PC}^1(\theta, [\theta_0, \infty)),$ if:
\begin{enumerate}
\item[(i)] it has discontinuities only at points $\theta_i, i \ge 0,$ and these  discontinuities are of the first kind; 
\item[(ii)] the derivative $z'(t)$ exists at each point $t \in  [\theta_0, \infty)\backslash \{\theta_i\},$ and 
at points $\theta_i, i \ge 0,$ the left-sided derivative exists;
\end{enumerate}

A solution $z(t)$ of (\ref{1+}) on $ [\theta_0, \infty)$ is a function $z(t) \in {\cal PC}^1(\theta, [\theta_0, \infty))$ such that:
\begin{enumerate}
\item[(1)] the differential equation  is satisfied for $z(t)$ on $ [\theta_0, \infty)\backslash \{\theta_i\},$ and it holds for the  left derivative of $z(t)$ at every point $\theta_i \ge 0.$ 
\item[(2)] the jumps equation  is satisfied by $z(t)$ for every   $ i \ge 0.$ 
\end{enumerate}

 Let us  denote $Z(t,s)$ the transition matrix  of the associated with (\ref{1+}) linear homogeneous system 
 \begin{eqnarray}
	&&z'(t) = A z(t), t \not = \theta_i\nonumber\\
	&&\Delta z|_{t = \theta_i} = B z(\theta_i).
	\label{1++}
\end{eqnarray}

 We may assume that:
 \begin{itemize}
	\item [(C5)] the matrices $A$ and $B$ commute, the real parts of all eigenvalues of the matrix 
	$A + \ln({\cal I} +B)$ are negative.
\end{itemize}
Condition $(C5)$ implies, \cite{samo}, Theorem 34,  that there exist two positive numbers  $N$ and $ \omega,$ which do not depend on $\theta,$  
\begin{eqnarray}
&&	||Z(t,s)|| \le N {\rm e}^{- \omega(t-s)} , t \ge s.
\label{exp}
\end{eqnarray}

Obviously, that condition for the matrices to commute is strong. One can obtain the exponential decay of the transition matrix using other conditions on matrices $A$ and $B,$ but  this would lengthen the paper significantly, and, moreover,  decrease the clarity of  the proof.

Assume additionally that 
\begin{itemize}	
\item [(C6)]$  NL	[\frac{2}{\omega}
	+ \frac{{\rm e}^{\omega}}{1 -{\rm e}^{-\omega}}] <1.$ 
\end{itemize}
The proof of following two  Theorems \ref{theorbd1} and \ref{theoremattr}
replicate those  of Theorems $37$ and  $89$ from \cite{samo}, respectively. 
   
 \begin{theorem} \label{theorbd1} If conditions $(C1),(C2),(C4) -(C6)$ are valid, then:
 
\begin{enumerate}
	\item   there exists a unique  bounded solution $z(t)$ of (\ref{1+}) from ${\cal PC}^1(\theta, \mathbb R),$
	
	\[
\]
	
 and $\| z(t)\| < NM_0[\frac{1}{\omega}+ \frac{{\rm e}^{\omega}}{1 -{\rm e}^{-\omega}}]$ for all $t \in \mathbb R;$
 
 \item  if there exists a number $p \in \mathbb N,$ such that  $\theta_{i+p} = \theta_i + p,  i \in \mathbb Z,$  then the bounded solution  $z(t)$ has the period $p;$ 
\end{enumerate} 
\end{theorem}

We may assume that
\begin{itemize}
	\item [(C7)]$ - \omega +NL + \ln(1+NL) < 0.$
\end{itemize}

Let $z(t)$ be the bounded solution, which exists by   Theorem \ref{theorbd1}, and  $z_1(t)$ be  another  solution of (\ref{1+}). 
The following theorem is valid. 

\begin{theorem} Assume that conditions $(C1),(C2),(C4) -(C7)$ are fulfilled. 

Then, for each $t,s \in \mathbb R, t \ge s,$ 
\begin{eqnarray} 
\|z(t) - z_1(t)\| \le \|z(s) - z_1(s) \|(1+NL){\rm e}^{(-\omega + NL + \ln(1+NL))(t-s)}. 
\label{exp23}
\end{eqnarray}
That is, the bounded solution $z(t)$ attracts all solutions of (\ref{1+}).
 \label{theoremattr}
\end{theorem}

In what follows we denote $z(t,\xi,v), \xi \in \mathbb R, v \in \mathbb R^n,$ a solution of  (\ref{1}) with $t_0 = \xi, z_0 =v,$ and assume that all conditions $(C1)-(C7)$ are fulfilled.

Assume that in (\ref{1+}) $\theta = \zeta(t_0), t_0 \in \Lambda.$ Then Theorem \ref{theorbd1} implies that for each $t_0 \in \Lambda$ there exists
a unique bounded on $\mathbb R$ solution of (\ref{1}). We denote the solution by $z(t,t_0).$ Moreover, if $\zeta(t_0)$ has the periodicity property with period $p,$ then the solution $z(t,t_0)$ is $p-$ periodic. Denote the periodic solutions  as $\phi(t,t_0).$  

 By Theorem  \ref{theoremattr} the bounded solution $z(t,t_0)$ attracts all solutions of  (\ref{1}) with the same points of discontinuity. That is, if $z(t,t_0,z_0)$ is another solution of (\ref{1}), then 

\begin{eqnarray} \label{exp24} \|z(t,t_0) - z(t,t_0,z_0)\| \le \|z(t_0,t_0) - z_0 \|(1+NL){\rm e}^{(-\omega + NL + \ln(1+NL))(t-t_0)}, 
\end{eqnarray}
for all $t \ge t_0.$ 
Finally, we remark that  there exist infinitely many periodic solutions $\phi(t,t_0), t_0 \in \Lambda.$  More precisely, for each $p \in \mathbb N$ there exists $t_0 \in \Lambda$ such that $z(t,t_0)$ is a $p-$periodic solution. These periodic solutions are different for diferent $p$ since  by condition $(C3)$ they have sequences of discontinuity points which do not intersect.

Denote ${\cal PC} = \{z(t,t_0,z_0):  t_0 \in \Lambda, z_0 \in \mathbb R^n\},$
and let ${\cal PCB} \subset {\cal PC}$ be the subset of all solutions bounded on $\mathbb R.$ That is, ${\cal PCB} = \{z(t,t_0):  t_0 \in \Lambda\}.$  On the basis of above made discussion we may say that the bounded set ${\cal PCB}$ is an attractor with the basin ${\cal PC}.$

\section{The chaos}

In this section we introduce the topological ingredients of the chaos for the dynamics of  (\ref{1})
and find the conditions for the existence of these ingredients. Moreover, we will prove that 
 ${\cal PCB}$  is the chaotic attractor, and discuss the period-doubling cascade for the problem.

\subsection{The chaotic attractor existence}

\begin{definition} We  say that (\ref{1}) is sensitive  on ${\cal PCB}$  if there exist  positive real numbers $\epsilon_0, \epsilon_1$ such that for each  $t_0 \in \Lambda,$  and 
for every $\delta > 0$ one could find a number  $t_1 \in \Lambda, |t_0 - t_1|  < \delta,$
and an interval  $Q$  from $[t_0, \infty)$ with length no less than $\epsilon_1$  such that
$||z(t,t_0) - z(t,t_1)|| \ge \epsilon_0, t \in Q,$ and there are no points of discontinuity of $z(t,t_0),z(t,t_1)$ in   $Q.$
 \label{defnsens}
\end{definition} 

To introduce following  definitions of density we need the concept of  "closeness"  for  piecewise continuous functions. Different types of metrics  and topologies for discontinuous functions are described in \cite{hw}, \cite{samo}-\cite{a4}, \cite{ak2},\cite{kol},\cite{skor}.  In what follows we are going to use the concept from \cite{kol}. 

Let us fix  $t_0, t_1 \in I$  and  an interval $J \subset  [t_0,\infty)\cap  [t_1,\infty),$ We  say that a function $\xi(t) \in {\cal PC}^1(\zeta(t_0),\mathbb R_{\zeta(t_0)})$ is $\epsilon-$equivalent to a function $\psi(t) \in {\cal PC}^1(\zeta(t_1),\mathbb R_{\zeta(t_1)})$
on $J$ and write $\xi(t)(\epsilon,J)\psi(t)$ if $\|\zeta(t_0) -\zeta(t_1)\|_J < \epsilon$ and 
$\| \xi(t)-\psi(t)\| < \epsilon$ for all $t$ from $J$ such that $t \not \in \cup_{\zeta_i(t_0),\zeta_i(t_1)\in J}\widehat{[\zeta_i(t_0),\zeta_i(t_1)]},$ where 
$\widehat{[a,b]}, a,b \in \mathbb R,$ means an oriented interval, that is $\widehat{[a,b]} = [a,b]$ if $a \le b,$ and
 $\widehat{[a,b]} = [b,a],$ if otherwise.  The equivalence of two piecewise continuous functions when  $\epsilon$ is small means, roughly speaking,  that they have close  discontinuity points, and the values of the functions are close at points that do not lie on  intervals between the neighbor discontinuity points of these functions. 

\begin{definition}  \label{defndensper} The set of all periodic  solutions  $\phi(t)=\phi(t,t_0), t_0 \in \Lambda,$ of (\ref{1}) is called dense in ${\cal PCB}$ if for every solution $ z(t,t_1), t_1 \in \Lambda,$   and each $\epsilon > 0,E  >0,$ there exists a periodic solution $\phi(t,t^*),t^* \in \Lambda, $ and an interval  $J \subset \mathbb R_{\zeta(t_1)}$ with length $E$ such that $\phi(t)(\epsilon,J)z(t).$ 
\end{definition} 
\begin{definition}  \label{defndense} A bounded solution $z(t) = z(t,t^*), t^* \in \Lambda,$ of  (\ref{1}) is called dense  in the set of all orbits of ${\cal PCB}$ if for every solution $z_1(t)= z(t,t_1), t_1 \in \Lambda,$ of  (\ref{1}),
 and each $\epsilon > 0,E  >0,$  there exists an interval  $J$ with length $E$  and a real number $\xi$ such that $z(t+ \xi)(\epsilon,J)z_1(t).$ 
\end{definition}

\begin{definition} \label{defnatrr}  The attractor ${\cal PCB}$ is chaotic if

\begin{enumerate}
	\item (\ref{1}) is sensitive in  ${\cal PCB};$
	\item the set of all periodic  solutions  $\phi(t,t_0), t_0 \in \Lambda,$ is dense in  ${\cal PCB};$
	\item  there exists a solution $z(t,t_0), t_0 \in \Lambda,$  which  is  dense  in ${\cal PCB}.$
\end{enumerate}
\end{definition}

\begin{theorem} Assume that conditions $(C1)-(C7)$ are fulfilled. Then the  set of all periodic  solutions  $\phi(t,t_0), t_0 \in \Lambda,$ of (\ref{1}) is  dense  in  ${\cal PCB}.$
 \label{theoremperiod}
\end{theorem}

\textbf{Proof:}  Let us fix $t_1 \in \Lambda$ and $E, \epsilon >0.$ By Lemma \ref{lempi} $(a)$ for  an arbitrary large number $\tilde T$  there exists a sequence $\zeta(t_0) \in \Pi$  with periodic property such that $\|\zeta(t_1) - \zeta(t_0)\|_Q < \epsilon,$ where $Q = (t_1,t_1 + \tilde T + E).$  We shall find the number  $\tilde T$ sufficiently large such that the bounded solution $z(t) = z(t,t_1)$ is $\epsilon-$equivalent to $\phi(t,t_0)$ on $J=(t_1 + \tilde T,t_1 + \tilde T + E).$
  Denote $M_1 =  NM_0(\frac{1}{\omega}+\frac{{\rm e}^{\omega}}{1 -{\rm e}^{-\omega}}).$ 

Let also  $Z_1(t,s)= Z(t,s,\zeta(t_1))$ and $Z_2(t,s)= Z(t,s,\zeta(t_0)), t \ge s,$ be the transition matrices. 
We have that \cite{samo}

$z(t) = Z_1(t,1) z(1) + \int\limits_{1}^t Z_1(t,s) f(z(s))ds + \sum\limits_{1 \le \zeta_i <t}Z_1(t,\zeta_i(t_1))W(z(\zeta_i(t_1))),$

$\phi(t) = Z_2(t,1)\phi(1) + \int\limits_{1}^t Z_2(t,s) f(\phi(s))ds + \sum\limits_{1 \le \zeta_i <t}Z_2(t,\zeta_i(t_0))W(\phi(\zeta_i(t_0))).$

It is difficult to  evaluate the difference between $z(t)$ and $\phi(t)$ using the last two expressions since
the moments of discontinuities do not coincide. For this reason let us apply the method of $B-$equivalence developed in papers \cite{a1}-\cite{a4}. 

Let us consider the following system
\begin{eqnarray}
	&&v'(t) = A v(t) + f(v), t \not  = \zeta_i(t_0),\nonumber\\
	&&\Delta v|_{t = \zeta_i(t_0)} = B v(\zeta_i(t_0)) + W(v(\zeta_i(t_0))) + W_i^1(v(\zeta_i(t_0))),
	\label{chop}
\end{eqnarray}

and the system 
\begin{eqnarray}
	&&z'(t) = A z(t) + f(z), t \not  = \zeta_i(t_1), \nonumber\\
	&&\Delta|_{t = \zeta_i(t_1)} = B z(\zeta_i(t_1)) + W(\zeta_i(t_1)),
	\label{chop1}
\end{eqnarray}
where $t_0,t_1$ are the numbers under discussion.  Assuming,  without loss of generality, that $\zeta_j(t_0)\le \zeta_j(t_1)$ for all $j,$  we introduce  

\[W_i^1(z) = ({\cal I} +B)\Big[( {\rm e}^{A(\zeta_i(t_1)-\zeta_j(t_0))}-{\cal I})z + \int_{\zeta_j(t_0)}^{\zeta_i(t_1)} {\rm e}^{A(\zeta_i(t_1)-s)}f(z(s))ds\Big] + \]\[
W( ({\cal I} +B)[{\rm e}^{A(\zeta_i(t_1)-\zeta_j(t_0))}z + \int_{\zeta_j(t_0)}^{\zeta_i(t_1)} {\rm e}^{A(\zeta_i(t_1)-s)}f(z(s))ds]) -  \]\[ \int_{\zeta_j(t_0)}^{\zeta_i(t_1)} {\rm e}^{A(\zeta_i(t_1)-s)}f(z_1(s))ds-  W(z),\]
where $z(t),z_1(t)$ are the solutions of 
\begin{eqnarray}
	&&z'(t) = A z(t)
	\label{chop2}
\end{eqnarray}
such that $z(\zeta_i(t_0))  = z$ and $z_1(\zeta_i(t_1)) = z(\zeta_i(t_1)+).$ One can easily verify that  $M_2 = \sup_{\|z\| \le M_1, i \in \mathbb Z}\|W_i^1(z)\| < \infty.$
Systems (\ref{chop}) and (\ref{chop1}) are $B-$equivalent \cite{a1}-\cite{a4}. That is their two solutions with the same initial data
coincide in their common domain if only  $t \not \in  \widehat{(\zeta_i(t_0),\zeta_i(t_1)]}, i \in \mathbb Z.$ 

So, if $v(t), v(1) = z(1),$ is the solution of (\ref{chop}), then $v(t)= z(t)$ for all $t \not \in  \widehat{(\zeta_i(t_0),\zeta_i(t_1)]}, i \in \mathbb Z.$

For  $v(t)$ we have that $$v(t) = Z_2(t,1)v(1) + \int\limits_{1}^t Z_2(t,s) f(v(s))ds +$$$$ \sum\limits_{1 \le \zeta_i <t}Z_2(t,\zeta_i(t_0))[W(v(\zeta_i(t_0))+W_1(v(\zeta_i(t_0))]. $$

 Consequently,
	\[\|\phi(t) - v(t)\| \le \|\phi(1) - v(1)\|  \|Z_2(t,1)\|+ 
	\int_{1}^t\|Z_2(t,s)\| L \|\phi(s) - v(s)\|ds  +  \]\[ \sum_{1 \le \zeta_j(t_0) <t}\|Z_2(t,\zeta_j(t_0))\|L\|\phi(\zeta_j(t_0)) - v(\zeta_j(t_0))\|  +   \sum_{1 \le \zeta_j(t_0) <t}\|Z_2(t,\zeta_j(t_0))\|\|W_1(v(\zeta_i(t_0))\| \le \]\[2M_1 N + M_2 \frac{{\rm e}^{\omega}}{1 -{\rm e}^{-\omega}}+	\int_{1}^t N{\rm e}^{-\omega(t- s)}L\|z(s) - v(s)\|ds + \]\[ \sum_{1 \le \zeta_j <t}N{\rm e}^{-\omega(t- \zeta_j(t_0))}L\|v(\zeta_j(t_0)) - v(\zeta_j(t_0))\|.	\]

 Now, applying the analogue of Gronwall-Bellman Lemma \cite{samo} for discontinuous functions, one can find that
 
\begin{eqnarray}
&&	\|z(t) - v(t)\| \le (2M_1 N + M_2 \frac{{\rm e}^{\omega}}{1 -{\rm e}^{-\omega}}) {\rm e}^{(-\omega + NL)(t- 1 )}\prod_{1 \le \zeta_j <t}(1+NL)
	 \le \nonumber\\
	 && (2M_1 N + M_2 \frac{{\rm e}^{\omega}}{1 -{\rm e}^{-\omega}}){\rm e}^{(-\omega + NL + \ln(1+NL))(t- 1 )}.	
\label{ho-ho}
\end{eqnarray}

Inequality (\ref{ho-ho}) implies that 
$\|z(t) - v(t)\| < \epsilon$ if $t > \tilde T,t \not \in  \widehat{[\zeta_i(t_0),\zeta_i(t_1)]}, i \in \mathbb Z,$ where $$ \tilde T = 1 + \frac{\ln(\frac{\epsilon}{2M_1 N + M_2 {\rm e}^{\omega}(1 -{\rm e}^{-\omega})^{-1}})}{-\omega + NL + \ln(1+NL)},$$
(we may assume that $\epsilon  < 2M_1$ ).
That is why, $z(t)(\epsilon,J)\phi(t)$ if  $J = (t_1+ \tilde T ,t_1+ \tilde T + E).$  
The theorem is proved.

\begin{theorem} Assume that conditions $(C1)-(C7)$ are fulfilled.  Then there exists a solution of (\ref{1})  $z(t,t^*)$ dense in  ${\cal PCB}.$
\label{theoremdense}
\end{theorem}
\textbf{Proof:} Fix positive $E,\epsilon.$  By Lemma \ref{lempi} $(b)$ there exists $t^* \in \Lambda$ such that $\zeta(t^*)$ is dense in $\Pi.$	 Denote $z_*(t) = z(t,t^*).$ 
Let us prove that $z_*(t)$ is the dense solution.

Consider an arbitrary solution $z(t) = z(t,t_0)\in {\cal PCB}.$    Consider an interval 
$J_1= (0,  E_1),$  where $E_1$ is an arbitrarily large positive number.  By Lemma \ref{lempi} $(b)$ there exists a natural $m$ such that  
\begin{eqnarray}
	&& \|\zeta(t_1) - \zeta(t^*,m)\|_{J_1} < \epsilon. 
	\label{105}
\end{eqnarray}
   We have that 
\[z_*(t+m) = {\rm e}^{A(t+m-1-m)}z_*(1+m) + \int_{1+m}^{t+m} {\rm e}^{A(t+m-u)}f(z_*(u))du +\]\[  \sum_{1+m \le \zeta_i(t_0) <t+m}  {\rm e}^{A(t+m- \zeta_i(t_0))}W(z_*(\zeta_i(t_0))) ={\rm e}^{A(t-1)}z_*(1+m) +  \]\[ \int_{1}^{t} {\rm e}^{A(t-u)}f(z_*(u+m))du + \sum_{1+m \le \zeta_i(t_0) <t+m}  {\rm e}^{A(t+m- \zeta_i(t_0))}W(z_*(\zeta_i(t_0))) .
\]
and 
\[z_1(t) = {\rm e}^{A(t-1)}z_1(1) + \int_{1}^{t} {\rm e}^{A(t-u)}f(z_1(u))du  + \sum_{s \le \zeta_i(t_1) <t}  {\rm e}^{A(t- \zeta_i(t_1))}W(z_1(\zeta_i(t_1))).
\]
	
   Now, using the last two formulas, similarly to proof of Theorem \ref{theoremperiod}, using (\ref{105}) and the   $B-$equivalence technique,  we can find
   a sufficiently large number $E_1 > 2E,$  and a natural number $m$ such that   $z_*(t+m)$ and $z_1(t)$ are $\epsilon-$equivalent on $J = (E_1/2, E_1).$ 
The theorem is proved.

Denote ${\overline m}= \max_{|u| \le 1}\|{\rm e}^{Au}\|, {\underline m}= \min_{|u| \le 1}\|{\rm e}^{Au}\|.$ 

Condition $(C3)$ implies that
	\[  \eta =\min_{\|x\| \le M_1}(Bx + W(x))> 0.
\]

Fix a number $q \ge 3,$ such that $\frac{1}{q} <\frac{2{\underline m}}{3{\overline m}}.$ 

We shall need the following assumption.
 \begin{itemize}
		\item [(C8)]$ L < \frac{[\frac{2{\underline m}}{3{\overline m}}-\frac{1}{q}]{\underline m} \eta}{2M_1({\overline m}+{\underline m})}.$	
\end{itemize}
 
\begin{theorem} Assume that conditions $(C1)-(C8)$ are fulfilled.  Then  (\ref{1}) is sensitive on ${\cal PCB}.$
\label{theoremsense}
\end{theorem}
\textbf{Proof:}  Fix a solution  $z(t) = z(t,t_0), t_0 \in \Lambda,$  and a positive 
 $\delta.$ 

Let $S(t_0) = s^0 = (s_0^0,s_1^0\ldots).$ Fix a number $t_1 \in \Lambda$ such that $S(t_1) = s^1 = (s_0^0,s_1^0\ldots,s_{\tilde n-1}^0,s_{\tilde n}^1,s_{\tilde n+1}^0,s_{\tilde n+2}^0,\ldots),s_{\tilde n}^1\not =s_{\tilde n}^0,$  for some $\tilde n > 0.$
We have that 
\begin{eqnarray*}
&&  d[\sigma^i s^0,\sigma^i s^1] = \left\{\begin{array}{rr} \frac{1}{2^{\tilde n-i}} & \mbox{if  $0 \le i \le \tilde n$},\\
0 & \mbox{if  $i >\tilde n$}.
\end{array}\right.
\end{eqnarray*}

Assume that $\tilde n \ge 3, \tilde n$ is  sufficiently large for $|t_0-t_1| = |S^{-1}(s^0) - S^{-1}(s^1)| < \delta.$

Now, denote 
$z_1(t) = z(t,t_1)$  the  solution of (\ref{1}).

Since $S$ is a homeomorphism and set $\Sigma_2$ is   compact, for a given $i,0 \le i \le \tilde n,$ the set 
	\[P_i = \{(\bar s,\tilde s)  \in \Sigma_2\times \Sigma_2: d[\bar s,\tilde s] \ge \frac{1}{2^{\tilde n-i}}\}
\]
is compact, and 
	\[\min_{(\bar s,\tilde s)\in P_i} |S^{-1}(\bar s) - S^{-1}(\tilde s)| = \mu_i >0,
\]
$P_{i+1} \subseteq P_i, \mu_{i+1} \ge \mu_i, 0 \le i < \tilde n.$
Fix $i_0 = \tilde n-2.$ Then  $|\kappa_i(t_0) - \kappa_i(t_1)| \ge \mu_{i_0}$ 
if $i = i_0, i_0+1.$

 Similarly, we also have  that  there exists a positive number $\mu_0 <1$ such that 
$|\kappa_i(t_0) - \kappa_i(t_1)| \le \mu_0$ if $0 \le i < \tilde n.$

Without loss of generality assume that  $\kappa_i(t_0) < \kappa_i(t_1)$ for all $i.$ Thus, there is a number $k$ among $ i_0, i_0+1$ such that 
$\kappa_k(t_1) - \kappa_k(t_0) > \mu_{i_0}$ and  $\kappa_k(t_0) - \kappa_{k-1}(t_1) \ge \frac{1}{2}(1- 
\mu_0).$

One can easily check that $(C8)$ implies that $\nu_1= \frac{2}{3} \frac{{\underline m}\eta}{\overline m} - 2LM_1 >0$
and $\nu_2 < \nu_1,$ where $\nu_2 =  \frac{2 {\overline m}LM_1}{{\underline m}} + \frac{1}{q} \eta.$

We shall show that  constants $\epsilon_0, \epsilon_1$ for Definition
\ref{defnsens}  can be taken equal to  $\epsilon_0 = \frac{1}{q} {\underline m}\eta, \epsilon_1 =\min\{\mu_{i_0},  \frac{1}{2}(1- 
\mu_0)\}.$

Assume that $\|z(\zeta_k(t_0)) - z_1(\zeta_k(t_0))\| < \nu_1.$ 

 Then, for $t \in [\zeta_k(t_0),\zeta_k(t_1)],$
 
	\[z(t) = {\rm e}^{A(t-\zeta_k(t_0))}({\cal I} +B)z((\zeta_k(t_0)) + \int_{\zeta_k(t_0)}^{t} {\rm e}^{A(t-s)}f(z(s))ds+   {\rm e}^{A(t-\zeta_k(t_0))}W(z((\zeta_k(t_0)) ),\]
	\[z_1(t) = {\rm e}^{A(t-\zeta_k(t_0))}z_1((\zeta_k(t_0)) + \int_{\zeta_k(t_0)}^{t} {\rm e}^{A(t-s)}f(z_1(s))ds.
\]
 We have that 
 
	\[\|z(t) - z_1(t)\| = \|{\rm e}^{A(t-\zeta_k(t_0))}[Bz((\zeta_k(t_0))+ W(z((\zeta_k(t_0)))] + \]\[{\rm e}^{A(t-\zeta_k(t_0))}[z((\zeta_k(t_0))- z_1((\zeta_k(t_0))] + \int_{\zeta_k(t_0)}^{t} {\rm e}^{A(t-s)}(f(z(s)) - f(z_1(s)))ds\| \ge\] \[  
{\underline m}\eta - {\overline m}( \nu_1 + 2LM_1) \ge \epsilon_0.\]

If $\|z(\zeta_k(t_0)) - z_1(\zeta_k(t_0))\| > \nu_2,$  then, for $t \in [\zeta_{k-1}(t_1),\zeta_k(t_0)],$
 
	\[z(t) = {\rm e}^{A(t-\zeta_k(t_0))}z((\zeta_k(t_0)) + \int_{\zeta_k(t_0)}^{t} {\rm e}^{A(t-s)}f(z(s))ds,\]
	\[z_1(t) = {\rm e}^{A(t-\zeta_k(t_0))}z_1((\zeta_k(t_0)) + \int_{\zeta_k(t_0)}^{t} {\rm e}^{A(t-s)}f(z_1(s))ds.
\]
and 
	\[\|z(t) - z_1(t)\| \ge {\underline m} \nu_2 - {\overline m}2LM_1  = \epsilon_0.\]
	The theorem is proved.
	
	On the basis of Theorems \ref{theoremperiod}-\ref{theoremsense} we can conclude that
	the following theorem is valid 
	
	\begin{theorem} \label{thmchaosattr} 	If conditions $(C1)-(C8)$ are fulfilled,  then ${\cal PCB}$ is a chaotic attractor.
	\end{theorem}

	Next, we are going to show that the attractiveness arranged by Theorem \ref{theoremattr}
	causes an amusing phenomena that the chaotic properties being attributed to all solutions of (\ref{1}), not only  those from ${\cal PCB}.$
	
	\subsection{The chaos of the initial value problem}
	
	Let us first introduce the following definitions.

\begin{definition} We  say that (\ref{1}) is sensitive  on $\Lambda$  if there exist  positive real numbers $\kappa_0, \kappa_1$ such that for every  solution $z(t) = z(t,t_0,z_0), t_0 \in \Lambda,$ of  (\ref{1}) and 
for each $\delta > 0, H >0,$ one could find a number  $t_1 \in \Lambda,$ such that $|t_0 - t_1| < \delta,$
and an interval  $Q$  from $[t_0, \infty)$ with length no less than $\kappa_1$  such that
$||z_1(t) - z(t)|| \ge \kappa_0, t \in Q,$  for all  solutions $z_1(t)=z(t,t_1,z_1), ||z_0 - z_1||  < H,$  of  (\ref{1}),   and there are no points of discontinuity of $z_1(t)$ and $z(t)$ on   $Q.$ 
 \label{defnsens1}
\end{definition} 
\begin{definition}  \label{defndensper1} The set of all periodic  solutions  $\phi(t,t_0), t_0 \in \Lambda,$ of (\ref{1}) is called dense in  ${\cal PC}$ if for every solution $z(t)= z(t,t_1,z_0), t_1 \in \Lambda, $ of the initial value problem, and each  $\epsilon >0, E >0,$ there exists a periodic solution $\phi(t)$ and an interval  $J \subset \mathbb R_{t_1}$ with length $E$ such that $\phi(t)(\epsilon,J)z(t).$ 
\end{definition} 
\begin{definition}  \label{defndense1} A solution $z(t) = z(t,t^*,z_0), t^* \in \Lambda, t \ge t^*,$ of  (\ref{1}) is called dense  in the set of all orbits of $\,{\cal PC}\,$ if for every solution $z_1(t) \in {\cal PC},$  and each  $\epsilon >0, E >0,$  there exists an interval  $J$ with length $E$  and a real number $\xi$ such that $z(t+ \xi)(\epsilon,J)z_1(t).$ 
\end{definition} 
\begin{definition}\label{defnchaos1}  Problem  (\ref{1}) is chaotic on $\Lambda$ if

\begin{enumerate}

  \item [(i)] it has the chaotic attractor ${\cal PCB};$ 
	\item [(ii)]it is sensitive on ${\cal PC};$
	\item [(iii)]the set of all periodic  solutions  $\phi(t,t_0), t_0 \in \Lambda,$ is dense in  ${\cal PC};$
	\item [(iv)] there exists a solution $z(t,t_0,z_0), t_0 \in \Lambda,$ of  (\ref{1}), which  is  dense  in the set of all orbits of  ${\cal PC};$
\end{enumerate}
\end{definition}

\begin{theorem} \label{mainthm} Assume that conditions $(C1)-(C8)$ are  fulfilled. Then (\ref{1}) is chaotic on $\Lambda.$
\end{theorem}
 \textbf{Proof:}  Condition  $(i)$ of Definition \ref{defnchaos1} is verified in Theorem \ref{thmchaosattr}.
 Using the attractiveness of ${\cal PCB}$ one can verify that parts  $(ii),(iii)$ and $(iv)$ are  also valid. Indeed, let us show first  that $(iv)$ is true. We will show that the solution $z_*(t) = z(t,t^*)$ dense in ${\cal PCB}$ by  Theorem \ref{theoremdense} is also dense in ${\cal PC}.$
Fix a solution $z(t)= z(t,t_0,z_0) \in {\cal PC}$ and a positive $\epsilon.$ 
The solution  $z(t)$ is attracted by the bounded solution $z(t,t_0) \in {\cal PCB}, z(t) \not = z(t,t_0),$ so that  (\ref{exp24}) is true.  Hence,  one can find a positive number $T$ such that $\|z(t,t_0,z_0) - z(t,t_0)\| < \epsilon/2$ if $ t \ge T.$ In the proof of Theorem \ref{theoremdense} it was shown that there exists an arbitrarily large  number $E_1,$  where we may assume that $E_1 > 2T,$ and a  natural number $m$ such that   $z_*(t+m)$ is $\epsilon/2-$equivalent to $z(t,t_0)$ on the  interval $J= (E_1/2,E_1).$ Now, it is easy to see that    $z_*(t+m)(\epsilon,J)z(t).$  So, $(iv)$ is prowen.  Part $(iii)$ can be verified in a similar way. Now consider condition  $(ii).$  To prove it, we will  follow the proof and the notation of Theorem \ref{theoremdense} and we will show that the  numbers in  Definition \ref{defnsens1} can be as $\kappa_0 = \epsilon_0/2$ and $\kappa_1= \epsilon_1,$ 
where $\epsilon_0$ and $\epsilon_1$ are the numbers used in the proof of Theorem \ref{theoremdense}.
 Consider a solution $z(t,t_0,z_0) \in {\cal PC}.$  From (\ref{exp24}) it follows that for a fixed $H>0$ there exists a number $T=T(H,z_0)$ such that  if $t \ge T,$ then  $\|z(t,\xi,z_1)-z(t,\xi)\| < \epsilon_0/4,$ where  $\xi \in \Lambda$ is arbitrary, and $\|z_1 - z_0\| <H.$ Analyzing the proof of Theorem \ref{theoremdense}
 one can readily observe  that the number $\tilde n,$ which satisfies all other demands of the proof, can be chosen greater than $T+2.$ Consequently,  the number $t_1 \in \Lambda$   corresponding to $\tilde n$  satisfies $|t_1-t_0| < \delta,$  and $\|z(t,t_0) - z(t,t_1)\| \ge \epsilon_0, t \in Q.$ Hence,
 $\|z(t,t_0,z_0)-z(t,t_1,z_1)\| \le \|z(t,t_0) - z(t,t_1)\|- \|z(t,t_0,z_0)-z(t,t_0)\|-\|z(t,t_1,z_1)-z(t,t_1)\| \ge \epsilon_0/2 = \kappa_0, t \in Q,$ where interval $Q$ has  length not less than $\kappa_1.$
 The theorem is proved.

\begin{remark} In \cite{lin} the author considers
the following initial value problem for a system of impulsive differential equations: 
\begin{eqnarray}
	&&x' = A(t)x +  G, t \not = \tau_i(m),\nonumber\\
	&&\Delta x|_{t = \tau_i(m)} =  I,\nonumber\\
	&&x(0) = x_0,
	\label{chop12}
\end{eqnarray}
where $t \in [0,\infty), x \in \mathbb R^n,$ the vector $x_0$ is fixed, $G$ and $I$ are  constant vectors, and 
\begin{itemize}
		\item [(D1)] the associated homogeneous system $x'= A(t)x$ is uniformly exponentially stable;
	\item [(D2)] an impulsive interval sequence $\tau_i(m), i = 1,2,\ldots,$ is defined  by the formula 
	
	\[\tau_i = \sum_{k=1}^i(m_k + \theta),  
\]
	where $m_k, k \ge 1,$ are iterations of 
	a sensitive map  $h$ defined on the bounded interval \, $\Xi \subset \mathbb R,$ such that there exists a number $\theta' >0$ satisfying $\theta + \inf\{t:t \in \Xi\} > \theta' >0.$
	\end{itemize} 
	The following definition was given in the paper.
\begin{definition}\label{++} \cite{lin} The  solution  of  (\ref{chop12})
 is said to be chaotic with respect to  the impulsive interval sequence  provided that $(i)$  this solution is uniformly bounded on the interval $[0,\infty);$  $(ii)$ there exist two positive numbers $\epsilon_0, \sigma >0,$ such that for 
 any $m \in \Xi$ one can find  a point $m^* \in \Xi$  arbitrarily close to $m,$  and a  positive number $T$ satisfying
 
	\[\mu (\{t_0 \le t \le T: \|x(t,m)-x(t,m^*)\| \ge \sigma\}) \ge \epsilon_0,
\]
 where $\mu(\cdot)$ represents the Lebesgue measure of a given set.
 \end{definition} 
 The author proves that under certain conditions, including  $(D1),(D2),$ the solution of  (\ref{chop12}) is chaotic in the sense of Definition \ref{++}.

Consider the following problem 
 \begin{eqnarray}
	&&x'(t) = 0, t \not = \tau_i(m),\nonumber\\
	&&\Delta x|_{t = \tau_i(m)} = (-1)^i,\nonumber\\
	&&x(0) = x_0,
	\label{chop13}
\end{eqnarray}
 
where  $t,x \in \mathbb R, x_0$ is fixed, and 
	\[\tau_i = \sum_{i=1}^i(m_i + 1),  m_1 = m, m \in (0,1), m_i = h(m_{i-1}), h(m) \equiv 0. 
\]
It is easily seen that $\tau_i = i + m-1, i = 1,2,\ldots$ Consider two solutions of the problem, $x(t,m), x(t,m^*), m,m^* \in (0,1), m \not = m^*.$ Obviously,
$\mu(\{0<t<k: |x(t,m)-x(t,m^*)| \ge 1\} \ge k|m-m^*|,$ where $k$ is a positive integer, and solution $x(t,m), m \in (0,1),$ is  bounded on $[0,\infty),$ as is  $x(t,m^*).$
In other words,  the solution is chaotic by Definition  \ref{++} even though  there is no "chaotic" assumption on the map $h,$ analogous to $(D2),$
and no condition similar to   $(D1)$ is imposed on the system.
 So, this example shows that a more scrupulous investigation of the problem is required. In particular, not only sensitivity, but also other ingredients  of chaos should be involved in the discussion.  We additionally emphasize that  sensitivity in \cite{lin} is defined  with respect to the   impulsive interval sequence, or, it is better to say, with respect to the parameter $m \in \Xi.$
Consequently, it cannot be considered a chaotic property.  Our definition of sensitiveness involves the initial data, and it is  natural  for dynamics \cite{lorenz, ruelle1, rob}.  Finally, we should remark that system (\ref{chop13}) is not sensitive in the sense of Definition \ref{defnsens1}.
\end{remark}
 
\subsection{The period-doubling cascade}

Now we consider $\mu >0$, $\mu$ being the parameter for the logistic map.

Since we use the logistic map as a tool to create the chaos, we expect to observe  the period-doubling cascade. 

Consider the logistic map  $h(t,\mu) \equiv  \mu t(1-t),$ assuming this time that $\mu < 4.$ 
It is known \cite{may}, that there exists an infinite sequence $3 <\mu_1< \mu_2<\ldots<\mu_k\ldots < 3.8284\ldots$ such that $h(t,\mu_i), i \ge 1,$ has an asymptotically stable    prime period$-2^i$ point $t_i^*$ with a region of attraction $(t_i^*-\delta_i, t_i^*+\delta_i).$ And beyond the value $3.8284\ldots,$ there are cycles with every integer period \cite{li}. 

By the results of our paper the period-doubling process is occurring in the bounded region $\|x\| < M_1$ of the space $\mathbb R^n,$ common for all $\mu >0.$  The chaotic properties  are observed in the region, and, by Theorem  \ref{theoremattr}, bounded solutions from this domain attract all other solutions. Moreover, the cascade generates infinitely many periodic solutions.
\begin{example} Consider the following initial value problem  
\begin{eqnarray}
	&&x_1'= 2/5x_2 + l\sin^2 x_2, \nonumber\\
	&&x_2'= 2/5x_1+ l\sin^2 x_1,	 t\not = \zeta_i(t_0), \nonumber\\
	&&\Delta x_1 |_{t = \zeta_i(t_0)} = -\frac{4}{3}x_1,\nonumber\\
	&&\Delta x_2 |_{t = \zeta_i(t_0)} = -\frac{4}{3}x_2 + W(x_2), 
	\label{ex3}
\end{eqnarray}
where   $W(s) = 1+s^2,$ if $|s| \le l,l$  is a positive constant,and $ W(s) = 1 + l^2,$ if $|s| > l.$ One can easily see that all the functions are lipschitzian with a constant proportional to $l.$ 
The matrices of coefficients are 

\begin{displaymath}
 A= \left ( \begin{array}{ccc}
0 \quad 2/5\\
2/5\quad 0 \end{array} \right),\quad B= \left ( \begin{array}{ccc}
-4/3 \quad\quad 0\\
\quad0\quad -4/3\end{array} \right).
\label{Amatrix}
\end{displaymath}
The matrices commute,  and the  eigenvalues of the matrix 
\begin{displaymath} A + {\mbox Ln}({\cal I}+ B) = \left ( \begin{array}{ccc}
-\ln 3 \quad\quad 2/5\\
\quad 2/5\quad -\ln 3 \end{array} \right)
\label{A+LnBmatrix}
\end{displaymath}
are  negative: $\lambda_{1,2} = -\ln 3 \pm 2/5 <0.$

Let us check if condition $(C3)$ holds.  It is sufficient to verify that 
 $ -\frac{1}{2}s + W(s) \not = 0,$ for all $s \in \mathbb R.$ 
 
 If $|s| \le l,$ then $-\frac{1}{2}s + W(s)= s^2-\frac{1}{2}s + 1,$ and it is never equal to zero.
 
 If $ |s| <l,$ then $-\frac{1}{2}s + W(s)= l^2-\frac{1}{2}x + 1.$ For the last expression to be zero, we need,  $s= -2(1+l^2),$ that is $|s| > l.$  We have a contradiction. Thus, condition $(C3)$ is valid. All the other conditions required by the Theorems  could be easily checked with sufficiently small coefficient $l.$  That is for $\mu > 4$ there is the chaotic behavior.
 
The numerical simulation of the chaos is not an easy task since even  the verification of  sensitivity requires two close values of the initial moment in the Cantor set $\Lambda,$ which cannot be found easily.   Hammel  {\it et all.} \cite{ham} have given a computer-assisted proof that an approximate trajectory of the logistic map can be shadowed  by  a true trajectory for a long time.  This result and the continuous dependence of the solutions on the sequence of  discontinuity points make possible the following appropriate simulations.

In what follows  we will demonstrate  for $\mu =4$ and $\mu = 3.8282$ the sensitivity and intermittency respectively.  
 
  Assume that $l = 10^{-2}.$ Choose $\mu = 4$ in (\ref{ex3})  and consider two solutions $x(t) = (x_1,x_2), \bar x(t)= (\bar x_1,\bar x_2),$ with  initial moments $t_0 = 7/9$ and $\bar t_0 = 7/9 + 3^{-12},$ respectively. That is, we take the initial values close to each other, and, moreover,   the solutions  with identical initial values, 
$x(t_0)= \bar x(\bar t_0) = (0.005,0.002).$
The graphs of the  coordinates of these solutions (Fig. 1) show that  the solutions abruptly become different  when $t$ is between $15$ and $20,$ despite being very close to each other for all $t$ in the interval $(\bar t_0,15).$  One can conclude that the phenomenon  of sensitivity  is numerically observable.

 Moreover, if one  consider  the sequence  $(x_1(n),x_2(n)), n = 1,2,3,\ldots,75000,$ in $x_1,x_2-$plane  (Fig. 2), then the chaotic attractor can be seen. 
 
\begin{figure}[hbp] 
\centering
\includegraphics[scale=0.5]{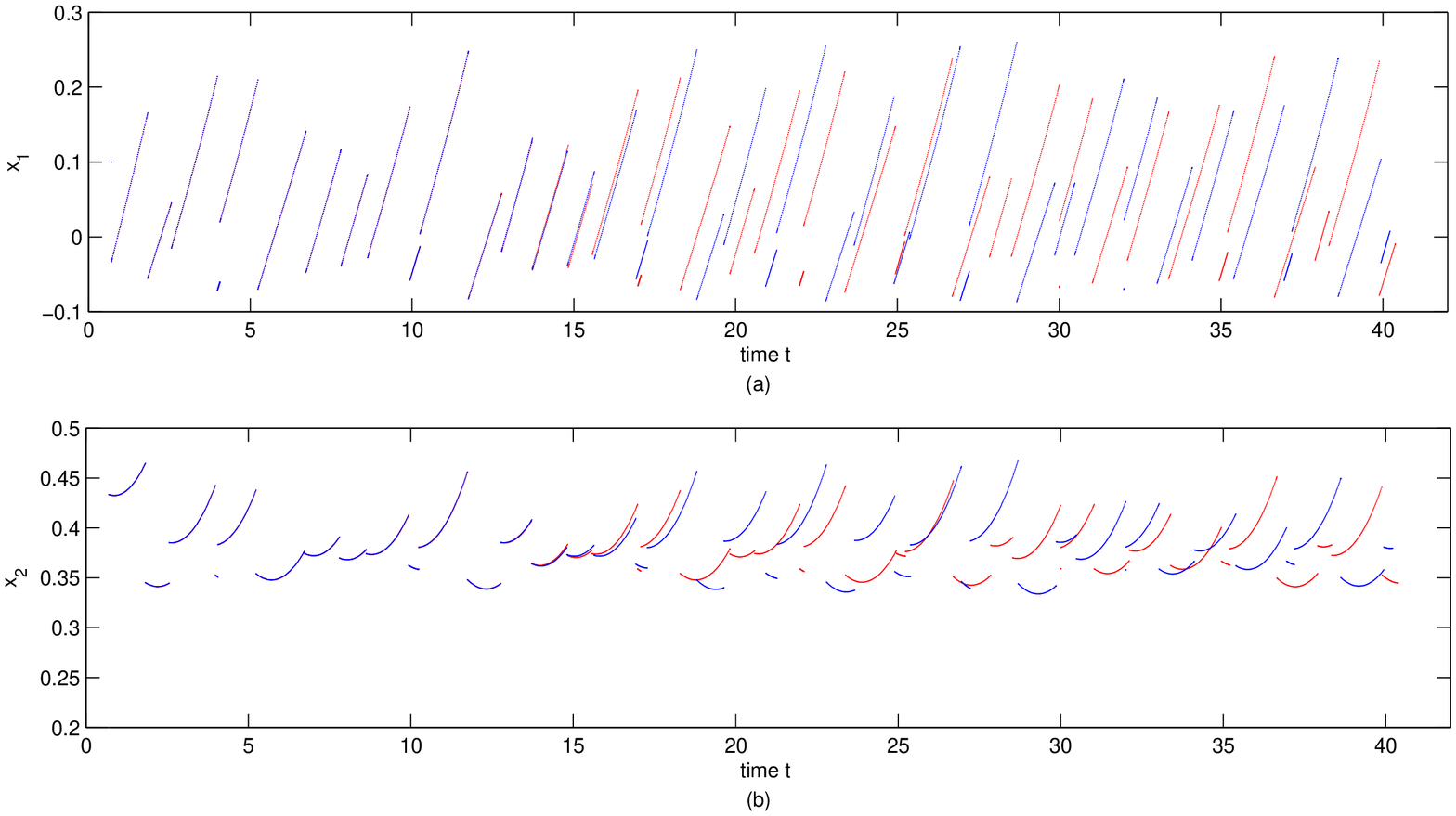}
\caption{Simulation results. Graphs of coordinates $x_1(t),x_2(t)$  are in the blue, and  of   coordinates $\bar x_1(t), \bar x_2(t)$ are red. The coordinates in $(a)$ and $(b)$ abruptly become significantly different when t is near $15$, while almost coinciding for all $t$ in the interval $(t_1,15).$  
}\label{fig1}
\end{figure}

\begin{figure}[hpbt]
 \begin{center}
  \epsfig {file=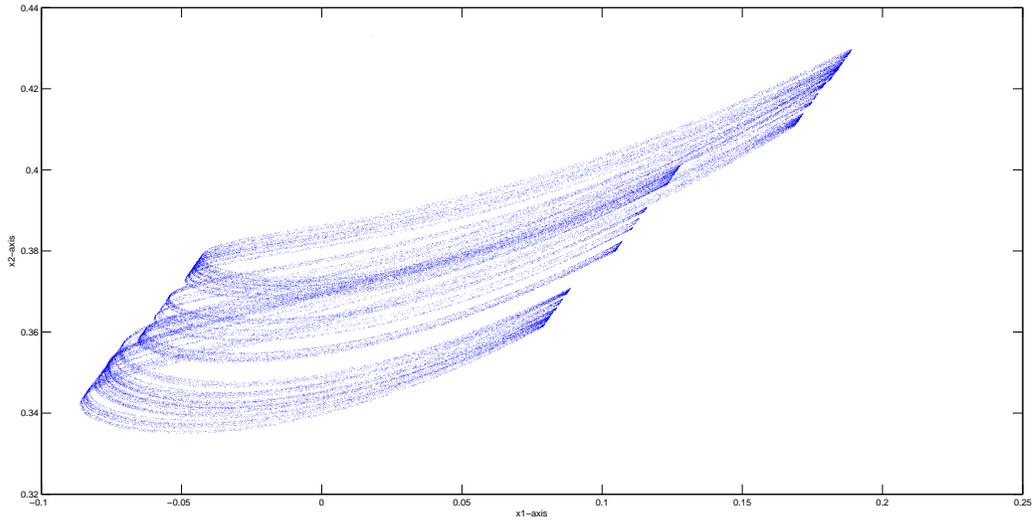,width=5.5in}
  \vspace{1.0cm}
  \caption{The chaotic attractor by a stroboscopic sequence $(x_1(n),x_2(n)), 1\le n \le 75000,$ is observable.}
\end{center}
\end{figure}

\begin{figure}[hbp] 
\centering
\includegraphics[scale=0.5]{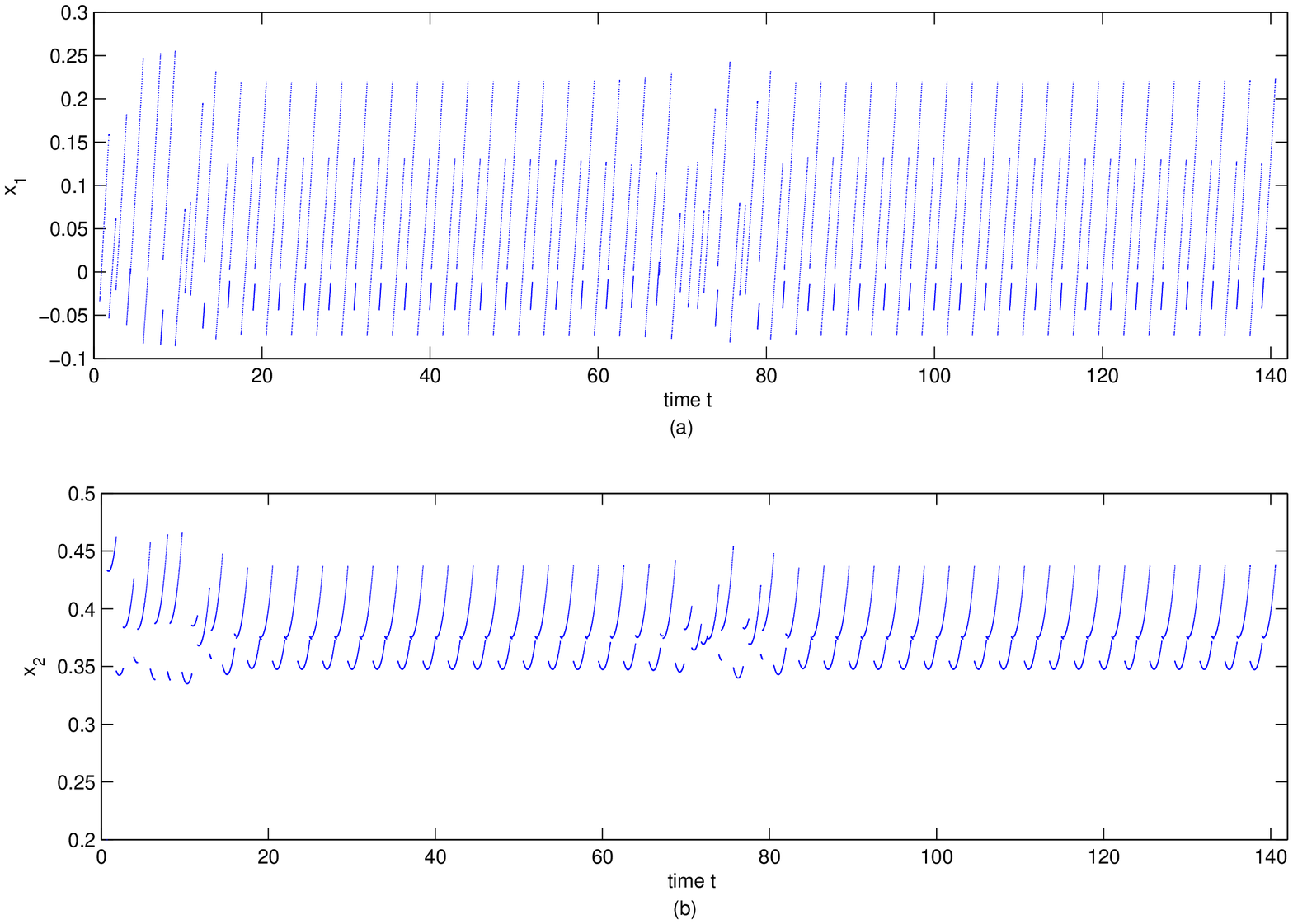}
\caption{Simulation results. In $(a)$ and $(b)$  for coordinates  $x_1$ and   $x_2$ respectively the intermittency phenomenon is indicated. 
}\label{fig2}
\end{figure}

Consider $\mu = 3.8282$  in (\ref{ex3}).   The coefficient's value of  $ 3.8282$ is such that the logistic map admits intermittency  \cite{str}.
Fix $t_0  = 0.5$ and take a solution $(x_1(t),x_2(t))$  of the last system  with the initial condition $x_1(t_0) = 0.002, x_2(t_0) = 0.005.$ The result of  simulation can be seen in Fig.2.

\end{example}
 
\section{Conclusion} 
The complex dynamics is obtained using Devaney's definition for guidance.
We also prove the existence of a chaotic attractor consisting of solutions bounded on the whole real axis, and observe the period-doubling cascade for the problem.
\begin{Ack}
This work was supported by Grant 106T418 from the Scientific and Technological Research Council of Turkey (TUBITAK). The author thanks  C. B\"{u}y\"{u}kadal\i \,  for the technical assistance. 

\end{Ack}  

\end{document}